\errorstopmode
\input amssym.def
\input amssym.tex


\magnification=\magstephalf
\hsize=14.0 true cm
\vsize=19 true cm
\hoffset=1.0 true cm
\voffset=2.0 true cm

\abovedisplayskip=12pt plus 3pt minus 3pt
\belowdisplayskip=12pt plus 3pt minus 3pt
\parindent=1.0em


\font\sixrm=cmr6
\font\eightrm=cmr8
\font\ninerm=cmr9

\font\sixi=cmmi6
\font\eighti=cmmi8
\font\ninei=cmmi9

\font\sixsy=cmsy6
\font\eightsy=cmsy8
\font\ninesy=cmsy9

\font\sixbf=cmbx6
\font\eightbf=cmbx8
\font\ninebf=cmbx9

\font\eightit=cmti8
\font\nineit=cmti9

\font\eightsl=cmsl8
\font\ninesl=cmsl9

\font\sixss=cmss8 at 8 true pt
\font\sevenss=cmss9 at 9 true pt
\font\eightss=cmss8
\font\niness=cmss9
\font\tenss=cmss10

\font\sixmib=cmmib6
\font\sevenmib=cmmib7
\font\eightmib=cmmib8
\font\ninemib=cmmib9
\font\tenmib=cmmib10

 at 12 true pt
 at 12 true pt
\font\bigrm=cmr10 at 12 true pt
 at 12 true pt
 at 12 true pt

 at 16 true pt
 at 16 true pt
\font\Bigrm=cmr12 at 16 true pt
 at 16 true pt
 at 16 true pt

\catcode`@=11
\newfam\ssfam
\newfam\mibfam

\def\tenpoint{\def\rm{\fam0\tenrm}%
    \textfont0=\tenrm \scriptfont0=\sevenrm \scriptscriptfont0=\fiverm
    \textfont1=\teni  \scriptfont1=\seveni  \scriptscriptfont1=\fivei
    \textfont2=\tensy \scriptfont2=\sevensy \scriptscriptfont2=\fivesy
    \textfont3=\tenex \scriptfont3=\tenex   \scriptscriptfont3=\tenex
    \textfont\itfam=\tenit                  \def\it{\fam\itfam\tenit}%
    \textfont\slfam=\tensl                  \def\sl{\fam\slfam\tensl}%
    \textfont\bffam=\tenbf \scriptfont\bffam=\sevenbf
                           \scriptscriptfont\bffam=\fivebf
                           \def\bf{\fam\bffam\tenbf}%
    \textfont\ssfam=\tenss \scriptfont\ssfam=\sevenss
                           \scriptscriptfont\ssfam=\sevenss
                           \def\ss{\fam\ssfam\tenss}%
    \textfont\mibfam=\tenmib \scriptfont\mibfam=\sevenmib
                             \scriptscriptfont\mibfam=\sevenmib
                             \def\mib{\fam\mibfam\tenmib}%
    \normalbaselineskip=13pt
    \setbox\strutbox=\hbox{\vrule height8.5pt depth3.5pt width0pt}%
    \let\big=\tenbig
    \normalbaselines\rm}

\def\ninepoint{\def\rm{\fam0\ninerm}%
    \textfont0=\ninerm      \scriptfont0=\sixrm
                            \scriptscriptfont0=\fiverm
    \textfont1=\ninei       \scriptfont1=\sixi
                            \scriptscriptfont1=\fivei
    \textfont2=\ninesy      \scriptfont2=\sixsy
                            \scriptscriptfont2=\fivesy
    \textfont3=\tenex       \scriptfont3=\tenex
                            \scriptscriptfont3=\tenex
    \textfont\itfam=\nineit \def\it{\fam\itfam\nineit}%
    \textfont\slfam=\ninesl \def\sl{\fam\slfam\ninesl}%
    \textfont\bffam=\ninebf \scriptfont\bffam=\sixbf
                            \scriptscriptfont\bffam=\fivebf
                            \def\bf{\fam\bffam\ninebf}%
    \textfont\ssfam=\niness \scriptfont\ssfam=\sixss
                            \scriptscriptfont\ssfam=\sixss
                            \def\ss{\fam\ssfam\niness}%
    \textfont\mibfam=\ninemib \scriptfont\mibfam=\sixmib
                            \scriptscriptfont\mibfam=\sixmib
                            \def\mib{\fam\mibfam\ninemib}%
    \normalbaselineskip=12pt
    \setbox\strutbox=\hbox{\vrule height8.0pt depth3.0pt width0pt}%
    \let\big=\ninebig
    \normalbaselines\rm}

\def\eightpoint{\def\rm{\fam0\eightrm}%
    \textfont0=\eightrm      \scriptfont0=\sixrm
                             \scriptscriptfont0=\fiverm
    \textfont1=\eighti       \scriptfont1=\sixi
                             \scriptscriptfont1=\fivei
    \textfont2=\eightsy      \scriptfont2=\sixsy
                             \scriptscriptfont2=\fivesy
    \textfont3=\tenex        \scriptfont3=\tenex
                             \scriptscriptfont3=\tenex
    \textfont\itfam=\eightit \def\it{\fam\itfam\eightit}%
    \textfont\slfam=\eightsl \def\sl{\fam\slfam\eightsl}%
    \textfont\bffam=\eightbf \scriptfont\bffam=\sixbf
                             \scriptscriptfont\bffam=\fivebf
                             \def\bf{\fam\bffam\eightbf}%
    \textfont\ssfam=\eightss \scriptfont\ssfam=\sixss
                             \scriptscriptfont\ssfam=\sixss
                             \def\ss{\fam\ssfam\eightss}%
    \textfont\mibfam=\eightmib \scriptfont\mibfam=\sixmib
                             \scriptscriptfont\mibfam=\sixmib
                             \def\mib{\fam\mibfam\eightmib}%
    \normalbaselineskip=10pt
    \setbox\strutbox=\hbox{\vrule height7.0pt depth2.0pt width0pt}%
    \let\big=\eightbig
    \normalbaselines\rm}

\def\tenbig#1{{\hbox{$\left#1\vbox to8.5pt{}\right.\n@space$}}}
\def\ninebig#1{{\hbox{$\textfont0=\tenrm\textfont2=\tensy
                       \left#1\vbox to7.25pt{}\right.\n@space$}}}
\def\eightbig#1{{\hbox{$\textfont0=\ninerm\textfont2=\ninesy
                       \left#1\vbox to6.5pt{}\right.\n@space$}}}

\font\sectionfont=cmbx10
\font\subsectionfont=cmti10

\def\figurecaptionfont{\ninepoint}
\def\tablecaptionfont{\ninepoint}
\def\footnotefont{\eightpoint}


\newcount\equationno
\newcount\bibitemno
\newcount\figureno
\newcount\tableno

\equationno=0
\bibitemno=0
\figureno=0
\tableno=0


\footline={\ifnum\pageno=0{\hfil}\else
{\hss\rm\the\pageno\hss}\fi}


\def\section #1. #2 \par
{\vskip0pt plus .10\vsize\penalty-100 \vskip0pt plus-.10\vsize
\vskip 1.6 true cm plus 0.2 true cm minus 0.2 true cm
\global\def\equationlabel{#1}
\global\equationno=0
\leftline{\sectionfont #1. #2}\par
\immediate\write\terminal{Section #1. #2}
\vskip 0.7 true cm plus 0.1 true cm minus 0.1 true cm
\noindent}


\def\subsection #1 \par
{\vskip0pt plus 0.8 true cm\penalty-50 \vskip0pt plus-0.8 true cm
\vskip2.5ex plus 0.1ex minus 0.1ex
\leftline{\subsectionfont #1}\par
\immediate\write\terminal{Subsection #1}
\vskip1.0ex plus 0.1ex minus 0.1ex
\noindent}


\def\appendix #1. #2 \par
{\vskip0pt plus .10\vsize\penalty-100 \vskip0pt plus-.10\vsize
\vskip 1.6 true cm plus 0.2 true cm minus 0.2 true cm
\global\def\equationlabel{\hbox{\rm#1}}
\global\equationno=0
\leftline{\sectionfont Appendix #1. #2}\par
\immediate\write\terminal{Appendix #1. #2}
\vskip 0.7 true cm plus 0.1 true cm minus 0.1 true cm
\noindent}



\def\equation#1{$$\displaylines{\qquad #1}$$}
\def\enum{\global\advance\equationno by 1
\hfill\llap{{\rm(\equationlabel.\the\equationno)}}}
\def\noenum{\hfill}
\def\next#1{\cr\noalign{\vskip#1}\qquad}
\def\nexteq#1{\cr\noalign{\vskip#1}\qquad}


\def\ifundefined#1{\expandafter\ifx\csname#1\endcsname\relax}

\def\ref#1{\ifundefined{#1}?\immediate\write\terminal{unknown reference
on page \the\pageno}\else\csname#1\endcsname\fi}

\newwrite\terminal
\newwrite\bibitemlist

\def\bibitem#1#2\par{\global\advance\bibitemno by 1
\immediate\write\bibitemlist{\string\def
\expandafter\string\csname#1\endcsname
{\the\bibitemno}}
\item{[\the\bibitemno]}#2\par}

\def\beginbibliography{
\vskip0pt plus .15\vsize\penalty-100 \vskip0pt plus-.15\vsize
\vskip 1.2 true cm plus 0.2 true cm minus 0.2 true cm
\leftline{\sectionfont References}\par
\immediate\write\terminal{References}
\immediate\openout\bibitemlist=biblist
\frenchspacing\parindent=1.8em
\vskip 0.5 true cm plus 0.1 true cm minus 0.1 true cm}

\def\endbibliography{
\immediate\closeout\bibitemlist
\nonfrenchspacing\parindent=1.0em}


\def\figurecaption#1{\global\advance\figureno by 1
\narrower\figurecaptionfont
Fig.~\the\figureno. #1}

\def\tablecaption#1{\global\advance\tableno by 1
\vbox to 0.25 true cm { }
\centerline{\tablecaptionfont%
Table~\the\tableno. #1}
\vskip-0.4 true cm}

\def\thintablerule{\hrule height0.4pt}

\tenpoint

\immediate\openin\bibitemlist=biblist
\ifeof\bibitemlist\immediate\closein\bibitemlist
\else\immediate\closein\bibitemlist

 \fi


\def\thismonth{\ifcase\month\or
January\or February\or March\or April\or May\or June\or
July\or August\or September\or October\or November\or December\fi}

\input epsf
\epsfclipon



\def\rmd{{\rm d}}

\def\rme{{\rm e}}
\def\rmO{{\rm O}}


\def\Im{{\rm Im}\,}
\def\Re{{\rm Re}\,}


\def\proof{\noindent{\sl Proof:}\kern0.6em}

\def\frac#1#2{\hbox{$#1\over#2$}}
\def\dual{\mathstrut^*\kern-0.1em}

\def\lvec#1{\setbox0=\hbox{$#1$}
    \setbox1=\hbox{$\scriptstyle\leftarrow$}
    #1\kern-\wd0\smash{
    \raise\ht0\hbox{$\raise1pt\hbox{$\scriptstyle\leftarrow$}$}}
    \kern-\wd1\kern\wd0}
\def\rvec#1{\setbox0=\hbox{$#1$}
    \setbox1=\hbox{$\scriptstyle\rightarrow$}
    #1\kern-\wd0\smash{
    \raise\ht0\hbox{$\raise1pt\hbox{$\scriptstyle\rightarrow$}$}}
    \kern-\wd1\kern\wd0}
\def\slash#1{\setbox0=\hbox{$#1$}\setbox1=\hbox{$\kern1pt/$}
    #1\kern-\wd0\kern1pt/\kern-\wd1\kern\wd0}


\def\nabstar#1{{\nabla\kern0.5pt\smash{\raise 4.5pt\hbox{$\ast$}}
               \kern-5.5pt_{#1}}}

\def\drvstar#1{{\partial\kern0.5pt\smash{\raise 4.5pt\hbox{$\ast$}}
               \kern-6.0pt_{#1}}}

\def\ldrvstar#1{{\lvec{\,\partial}\kern-0.5pt\smash{\raise 4.5pt\hbox{$\ast$}}
               \kern-5.0pt_{#1}}}


\def\MSbar{\overline{\rm MS\kern-0.5pt}\kern0.5pt}



\def\psibar{\overline{\psi}{\vphantom{\psi}}\kern-0.6pt}
\def\chibar{\overline{\chi}{\vphantom{\chi}}\kern-0.6pt}
\def\Vbar{\kern1.0pt\overline{\kern-1.0ptV\kern-1.0pt}\kern1.0pt{\vphantom{V}}}
\def\cbar{\bar{c}}
\def\ctilde{\tilde{c}{\kern1pt\vphantom{c}}}
\def\cbartilde{\tilde{\cbar}{\kern1pt\vphantom{c}}}
\def\dbar{\bar{d}}
\def\dtilde{\tilde{d}{\kern1pt\vphantom{d}}}
\def\dbartilde{\tilde{\dbar}{\kern1pt\vphantom{d}}}
\def\phit{\tilde{\phi}}


\def\diracstar#1#2{
    \setbox0=\hbox{$\gamma$}\setbox1=\hbox{$\gamma_{#1}$}
    \gamma_{#1}\kern-\wd1\kern\wd0
    \smash{\raise4.5pt\hbox{$\scriptstyle#2$}}}


\def\SUn{{\rm SU}(N)}


\def\om{\omega}

\def\eps{\epsilon}

%
\rightline{CERN-PH-TH/2011-035}
\vskip1.2cm 
\centerline{\Bigrm
Non-renormalizability of the HMC algorithm}
\vskip 0.6 true cm
\centerline{\bigrm Martin L\"uscher and Stefan Schaefer}
\vskip1.5ex
\centerline{{\it CERN, Physics Department, 1211 Geneva 23, Switzerland}}
\vskip 0.8 true cm
\thintablerule
\vskip 2.0ex
\ninepoint
\leftline{\bf Abstract}
\vskip 1.0ex\noindent
In lattice field theory, 
renormalizable simulation algorithms are attractive, because their
scaling behaviour as a function of the lattice spacing is predictable.
Algorithms implementing the Langevin equation, for example,
are known to be renormalizable if the simulated theory is.
In this paper we show that the situation is different
in the case of the molecular-dynamics evolution on which 
the HMC algorithm is based. 
More precisely, studying the $\phi^4$ theory, we find that
the hyperbolic character of the molecular-dynamics equations
leads to non-local (and thus non-removable) 
ultraviolet singularities already at one-loop order of 
perturbation theory. 
\vskip 2.0ex
\thintablerule

\tenpoint

\vskip-0.3cm

\section 1. Introduction

Numerical simulations in lattice field theory are based on
stochastic processes that produce random sequences of representative 
field configurations. It is often useful
to interpret the simulation time in these calculations 
as a further space-time coordinate.
The $n$-point autocorrelation functions of the local fields 
then formally look like the correlation functions in a field theory 
with an extra dimension and
they are, in fact, sometimes representable in this way.
Depending on the simulation algorithm, and 
if the simulated theory is renormalizable,
the autocorrelation functions may conceivably 
be renormalizable as well. 
The scaling properties of such 
algorithms (which, for brevity, will be referred 
to as renormalizable)
are encoded in the continuum theory and thus
become predictable to some extent.

In the pure $\SUn$ gauge theory, for example, 
simulation algorithms that integrate
the Langevin equation are known to be renormalizable
[\ref{Zinn},\ref{ZinnZwanziger}]. 
The integrated autocorrelation times $\tau_{\rm int}$ 
of physical observables have dimension
$[{\rm length}]^2$ in this case. Moreover,
the standard renormalization group analysis and
a one-loop calculation in perturbation theory 
[\ref{MunozAlvarez},\ref{Okano}] imply that they scale 
according to [\ref{BaulieuZwanziger}]
\equation{
  \tau_{\rm int}=Cg_0^{9/11}\left\{1+\rmO(g_0^2)\right\}r_0^2
  \enum
}
at small lattice spacings $a$, where $C$ is an observable-dependent constant,
$g_0$ the bare gauge coupling and $r_0$ the Sommer radius [\ref{SommerScale}].
In lattice units, the
autocorrelation times thus increase like $1/a^2$ as $a\to0$ up
to a logarithmically decreasing factor%
\kern1pt\footnote{$\dagger$}{\footnotefont%
Equation (1.1) is expected to hold on the infinite lattice and on
finite lattices with open boundary conditions [\ref{Villasimius}]. 
On lattices with periodic boundary conditions,
topology-changing tunneling transitions 
can give rise to very large autocorrelation times
[\ref{DelDebbioTauQ}--\ref{Villasimius}]. 
Such transitions are lattice artifacts and are 
therefore not included in the analysis that leads to eq.~(1.1).
}.

Most simulations of lattice QCD performed today are based on
some variant of the HMC algorithm [\ref{HMC}]. 
The form of the underlying molecular-dynamics equations and 
free-field studies [\ref{GHMC}] suggest that 
the simulation time has physical dimension $[\rm length]$ in 
this case and that the autocorrelation times consequently scale 
essentially like $1/a$. As far we know, the 
renormalizability of the algorithm has however never been 
studied and its scaling properties in presence
of interactions thus remain unknown.

In this paper, the issue is addressed in the 
framework of perturbation theory. For simplicity
the $\phi^4$ theory is considered, but our main result
(the non-renormalizability of the molecular-dynamics
equations) no doubt extends to most theories of interest.
A slightly generalized version of the HMC algorithm
is studied, which was introduced many years ago by 
Horowitz [\ref{Horowitz}] (see sects.~2 and 3). 
The non-renormalizability of the associated stochastic
equation is then established 
by showing that the four-point autocorrelation function of the 
fundamental field has a non-removable ultraviolet singularity 
at second order in the coupling (sects.~4 and 5).

\section 2. Stochastic molecular dynamics

In order to simplify the discussion as much as possible, 
we consider the $\phi^4$ theory with a single scalar field $\phi$
and dimensional instead of a lattice regularization. 
The action of the field in $D=4-2\eps$ Euclidean dimensions 
is given by
\equation{
  S=\int\rmd^D\kern-1pt x\,\left\{
  {1\over2}\partial_{\mu}\phi(x)\partial_{\mu}\phi(x)+
  {1\over2}m_0^2\phi(x)^2+
  {g_0\over4!}\phi(x)^4\right\},
  \enum
}
where $m_0$ denotes the bare mass parameter 
and $g_0$ the bare coupling constant. 
The generalized HMC algorithm [\ref{Horowitz}] 
integrates a stochastic version of
the molecular-dynamics equations that derive from the action (2.1).
In the following subsections, we briefly discuss
these equations and solve them in powers of
the coupling $g_0$.

\subsection 2.1 Evolution equations

As usual the molecular dynamics evolves 
the field $\phi(t,x)$ together with its momentum $\pi(t,x)$ as a 
function of a fictitious time $t$.
The stochastic evolution equations [\ref{Horowitz}]
\equation{
  \partial_t\phi=\pi,
  \enum
  \nexteq{3.0ex}
  \partial_t\pi=-{\delta S\over\delta\phi}-2\mu_0\pi+\eta
  \noenum
  \nexteq{2.0ex}
  {\phantom{\partial_t\pi}}
  =(\partial_{\mu}\partial_{\mu}-m_0^2)\phi-{g_0\over3!}\phi^3
  -2\mu_0\pi+\eta,
  \enum
}
involve another mass parameter, $\mu_0>0$, and a Gaussian noise 
$\eta(t,x)$ with vanishing expectation value and variance
\equation{
  \langle\eta(t,x)\eta(s,y)\rangle=4\mu_0\delta(t-s)\delta(x-y).
  \enum
}
Evidently, the ordinary molecular dynamics  
is recovered in the limit $\mu_0\to0$.
Moreover, in the second-order form,
\equation{
  \partial_t^2\phi+2\mu_0\partial_t\phi
  =-{\delta S\over\delta\phi}+\eta,
  \enum
}
and after substituting $t\to2\mu_0t$, the evolution equations
are seen to coincide with the Langevin equation up to a term 
that goes to zero at large $\mu_0$.

Since its introduction by Horowitz [\ref{Horowitz}],
the generalized HMC algorithm has been 
occasionally studied in the literature, where it is
referred to as the Kramers equation or the L2MC algorithm 
(see refs.~[\ref{BeccariaCurci},\ref{JansenLiu},\ref{GHMC}],
for example). 
In practice, one starts from the first-order equations
(2.2),(2.3) and implements the algorithm using
symplectic integrators and acceptance-rejection
steps. For the theoretical analysis in this paper,
we however prefer to proceed with the second-order equation (2.5).

\subsection 2.2 Solution of eq.~(2.5) to leading order in $g_0$

The leading-order equation
\equation{
   {\cal D}\phi_0=\eta,
   \enum
   \nexteq{2.5ex}
   {\cal D}=\partial_t^2+2\mu_0\partial_t-\partial_{\mu}\partial_{\mu}+m_0^2,
   \enum
}
coincides with the Klein--Gordon equation in 
$D+1$ dimensions except for the term proportional to $\mu_0$,
which tends to damp the time evolution of the field. 
At large $\mu_0$ and after a rescaling of $t$, 
the equation actually turns into the heat equation.

The Green function
\equation{
   K(t,x)=\int_{\om,p}\rme^{-i\om t+ipx}\tilde{K}(\om,p),
   \enum
   \nexteq{2.5ex}
   \tilde{K}(\om,p)=\left(-\om^2-2i\mu_0\om+p^2+m_0^2\right)^{-1},
   \enum
}
of the differential operator ${\cal D}$ is discussed in some detail 
in appendix A. Here and below, the notational convention
\equation{
   \int_{\om}=\int{\rmd\om\over 2\pi},
   \qquad
   \int_p=\int{\rmd^D\kern-1pt p\over(2\pi)^D},
   \enum
}
is used. It is then straightforward to show that the solution 
of the wave equation (2.6) at time $t\geq t_0$ with prescribed 
initial data at time $t_0$ is given by
\equation{
   \phi_0(t,x)=\int_{t_0}^t\rmd s\,\int\rmd^D\kern-1pt y\,K(t-s,x-y)\eta(s,y)
   \noenum
   \nexteq{1.5ex}
   \kern0.5em
   +\int\rmd^D\kern-1pt y\left\{K(t-t_0,x-y)(\partial_t\phi_0)(t_0,y)
   +(\partial_t+2\mu_0)K(t-t_0,x-y)\phi_0(t_0,y)\right\}.
   \noenum
   \nexteq{0.5ex}
   \enum
}
Note that the dependence on the initial data dies away exponentially 
with increasing time (see appendix A). The stochastic 
molecular-dynamics evolution thus thermalizes and eventually
loses all memory of the initial values of the field.

In the following, we
shall only be interested in the behaviour of the autocorrelation
functions after thermalization. We therefore move the thermalization
phase to time $t_0=-\infty$ and are then left with the solution
\equation{
   \phi_0(t,x)=\int_{-\infty}^t\rmd s\,
   \int\rmd^D\kern-1pt y\,K(t-s,x-y)\eta(s,y)
   \enum
}
of eq.~(2.6) that describes the stationary situation.

\subsection 2.3 Iterative solution of the evolution equation

Equation (2.5) may be written in the form
\equation{
   {\cal D}\phi=\eta-{g_0\over3!}\phi^3
   \enum
}
or, equivalently, as an integral equation
\equation{
   \phi(t,x)=\phi_0(t,x)
   -{g_0\over3!}\int_{-\infty}^t\rmd s\,\int\rmd^D\kern-1pt y\,
   K(t-s,x-y)\phi(s,y)^3.
   \enum
}
Iteration of the latter then yields the solution $\phi(t,x)$ in
powers of $g_0$.

\topinsert
\vbox{
\vskip0.0cm
\centerline{\epsfysize=2.5cm\epsfbox{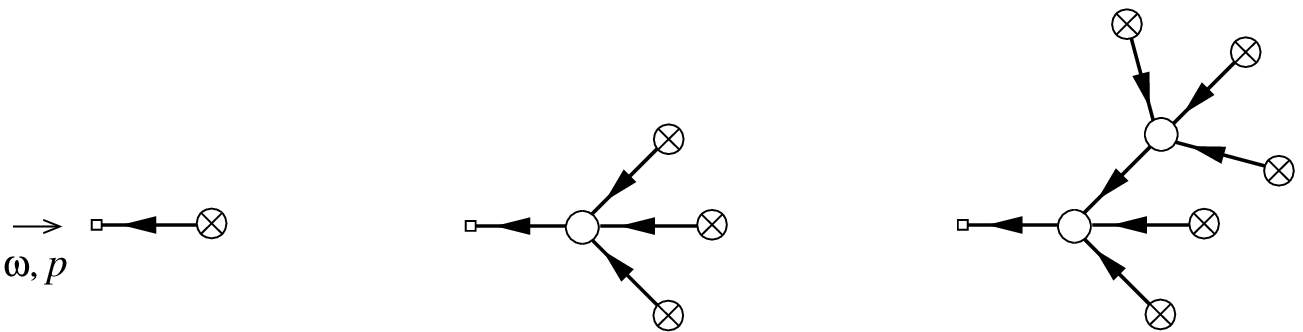}}
\vskip0.5cm
\figurecaption{%
In perturbation theory, the solution of the integral equation (2.14) 
is given by a series of directed tree diagrams. 
The diagrams up to second order in $g_0$ are shown in this figure.
All diagrams have a single external line (labeled by
a little square) with ingoing frequency-momentum
$(\om,p)$. The arrows on the internal lines all point
in the direction towards the external line.
}
}
\endinsert

Each term in this expansion may be represented by a tree diagram
with directed lines, four-point and one-point vertices
(see fig.~1). In frequency-momentum space,
\equation{
   \tilde{\phi}(\om,p)=
   \int\rmd t\,\rmd^D\kern-1pt x\,\rme^{i\om t-ipx}\phi(t,x),
   \enum
}
the lines represent the Green function
\equation{
  \raise-0.73cm\hbox{\epsfxsize=1.7cm\epsfbox{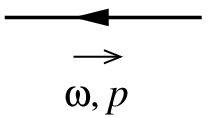}}%
  =\;\tilde{K}(\om,p),
  \enum
}
while the one-point vertices 
\equation{
  \raise-0.095cm\hbox{\epsfxsize=2.0cm\epsfbox{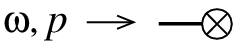}}%
  \;=\;\tilde{\eta}(\om,p)
  \enum
}
stand for the insertion of the noise field. 
As in ordinary Feynman diagrams, there is a 
frequency-momentum conservation $\delta$-function
\equation{
   (2\pi)^{D+1}
   \delta(\om_1+\om_2+\om_3+\om_4)
   \delta(p_1+p_2+p_3+p_4)
   \enum
}
associated to each vertex
\equation{
  \raise-0.39cm\hbox{\epsfxsize=1.0cm\epsfbox{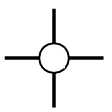}}%
  \;=\;-g_0
  \enum
}
with ingoing frequency-momenta $(\om_1,p_1),\ldots,(\om_4,p_4)$.
The lines in these diagrams are directed, because $\tilde{K}(\om,p)$ is
not invariant under a change of sign of $\om$. In position space,
the arrows point in the direction of increasing simulation time.

\section 3. Autocorrelation functions

The $n$-point autocorrelation functions of the field $\phi(t,x)$
are usually defined by taking the time average 
of the product $\phi(t_1,x_1)\ldots\phi(t_n,x_n)$ at fixed
time lags $t_i-t_j$.
In the present setup, the translation symmetry in time
allows the time average to be replaced by 
the average $\langle\ldots\rangle$ over the noise field $\eta(s,y)$.
We are thus led to consider the correlation functions
\equation{
   \tilde{\cal A}_n(\om_1,p_1;\ldots;\om_n,p_n)=
   \langle\phit(\om_1,p_1)\ldots\phit(\om_n,p_n)\rangle
   \enum
}
in frequency-momentum space,
which may be computed in perturbation theory by expanding the fields 
$\phit(\om_k,p_k)$ in powers of the coupling $g_0$, following 
the lines of the previous section, and by contracting 
the noise fields using Wick's rule. 
As a result one obtains a sum of Feynman diagrams 
for the autocorrelation functions
similar to the ones for the ordinary 
(field-theoretical) correlation functions.

\subsection 3.1 Feynman rules

The one-point vertices
in the tree diagrams that represent the terms in the
expansion of $\phit(\om_k,p_k)$
are connected to the rest of the tree through a directed line.
When the noise fields at any two such vertices are contracted, 
an undirected line
\equation{
  \raise-0.77cm\hbox{\epsfxsize=1.8cm\epsfbox{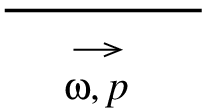}}%
  =\;\tilde{G}(\om,p)
  \enum
}
is obtained, where
\equation{
   \tilde{G}(\om,p)=4\mu_0\tilde{K}(\om,p)\tilde{K}(-\om,p).
   \enum
}
In the case of the two-point autocorrelation function, for example,
the contraction of the lowest-order diagram
\equation{
  \raise-0.5cm\hbox{\epsfxsize=4.2cm\epsfbox{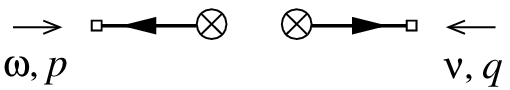}}%
  \enum
}
leads to
\equation{
   \langle\tilde{\phi}(\om,p)\tilde{\phi}(\nu,q)\rangle
   =(2\pi)^{D+1}\delta(\om+\nu)\delta(p+q)\tilde{G}(\om,p)
   +\rmO(g_0).
   \enum
}
The Feynman diagrams for the autocorrelation functions thus
involve directed lines (2.16), undirected lines (3.2) and the 
vertices (2.19).
Given these lines and vertices, the Feynman rules are the usual 
ones except for the following special features:

\vskip1ex\noindent
(a) Both kinds of lines (directed and undirected) can connect
to the vertices. The value of the latter is the same in all cases. 

\vskip1ex\noindent
(b) Each vertex has exactly one outward-directed
line attached to it. This line may be an internal or an external line.

\vskip1ex\noindent
(c) There are no diagrams with loops of directed lines. 

\vskip1ex\noindent
(d) External lines may be outward-directed lines or undirected lines. 
There are no inward-directed external lines. 

\vskip1ex\noindent
The structure of the Feynman diagrams is otherwise the same as in
any field theory. In particular, the diagrams can be decomposed into
one-particle irreducible parts and the lines connecting them.
Note that the external legs of the irreducible parts 
may be directed or undirected (see fig.~2). The
two types of legs must be distinguished and there
are thus many more irreducible diagrams than in the 
case of the ordinary correlation functions.

\topinsert
\vbox{
\vskip0.0cm
\centerline{\epsfxsize=8.0cm\epsfbox{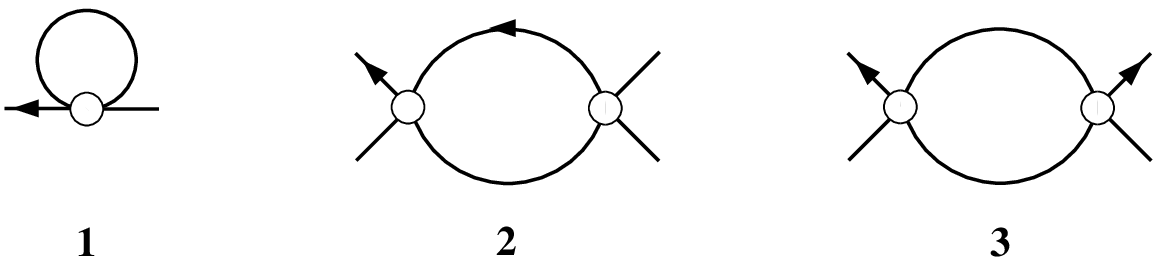}}
\vskip0.4cm
\figurecaption{%
One-loop vertex diagrams contributing to the two- and four-point
correlation functions of the basic field. Up to permutations
of the external lines, these are all 
one-loop vertex diagrams with less than six legs.
}
}
\endinsert

\subsection 3.2 Computation of the two-point function to one-loop order

The decomposition of the two-point autocorrelation function
into one-particle irreducible parts reads
\equation{
   \tilde{\cal A}_2(\om,p;\nu,q)=(2\pi)^{D+1}\delta(\om+\nu)\delta(p+q)
   \bigl\{G(\om,p)
   \noenum
   \next{2.0ex}
   \qquad
   +K(\om,p)\Sigma(\om,p)G(\om,p)+
   G(\om,p)\Sigma(-\om,p)K(-\om,p)+\ldots\bigr\},
   \enum
}
where the self-energy $\Sigma$ is given 
to one-loop order by the diagram 1 in fig.~2.
A short calculation, using eq.~(A.7), then shows that
\equation{
   \Sigma(\om,p)=-{g_0\over2}I_1+\rmO(g_0^2),
   \enum
   \nexteq{2.5ex}
   I_1=\int_q(q^2+m_0^2)^{-1}
   \mathrel{\mathop=_{\eps\to0}}
   -{m_0^2\over16\pi^2\eps}+\rmO(1).
   \enum
}
To this order, the self-energy thus
coincides with the familiar tadpole diagram
that contributes to the ordinary two-point correlation function.
In particular, the poles at $\eps=0$ will later be seen 
to cancel when the mass parameter $m_0$ is renormalized.

\subsection 3.3 Computation of the diagrams 2 and 3

Apart from diagram 1 
(which may be inserted in the 
external lines), 
the diagrams 2 and 3 in fig.~2 are the only
one-particle irreducible diagrams that contribute to the four-point
autocorrelation function at one-loop order.
Up to a factor $g_0^2$ and the statistical factor,
they are given by 
\equation{
   I_2(\om,p)=\int_{\nu,q}\tilde{K}(\nu+\om,q+p)\tilde{G}(\nu,q),
   \enum
   \nexteq{2.5ex}
   I_3(\om,p)=\int_{\nu,q}\tilde{G}(\nu+\om,q+p)\tilde{G}(\nu,q),
   \enum
}
where $(\om,p)$ is the external frequency-momentum that flows
through the diagrams from left to right.

The integrals $I_2$ and $I_3$ can be transformed to a
useful alternative form by inserting
the time-momentum representation of the propagators
(see appendix A). Taking the support properties of 
the Green function into account, one first notes that
\equation{
   I_2(\om,p)=
   -{1\over2}\int_q\int_0^{\infty}\rmd t\,
   \rme^{i\om t}\partial_t\bigl\{
   \hat{G}(t,q+p)\hat{G}(t,q)\bigr\}.
   \enum
}
Partial integration then yields
\equation{
   I_2(\om,p)={1\over2}J_0(p)+{i\om\over2}J_1(\om,p),
   \enum
}
where
\equation{
   J_0(p)=\int_q\left\{((q+p)^2+m_0^2)(q^2+m_0^2)\right\}^{-1},
   \enum
   \nexteq{2.5ex}
   J_1(\om,p)=\int_q\int_0^{\infty}\rmd t\,\rme^{i\om t}
   \hat{G}(t,q+p)\hat{G}(t,q).
   \enum
}
The integral $I_3$ is similarly given by 
\equation{
   I_3(\om,p)=J_1(\om,p)+J_1(-\om,p)
   \enum
}
so that it suffices to work out the integrals $J_0$ and $J_1$.

The integral $J_0$ coincides with the familiar one-loop diagram
contributing to the ordinary four-point correlation function in 
the $\phi^4$ theory. 
It is logarithmically divergent and thus has a pole,
\equation{
   J_0(p)\mathrel{\mathop=_{\eps\to0}}
   {1\over16\pi^2\eps}+\rmO(1),
   \enum
}
at $\eps=0$. In four dimensions,
the integral $J_1$ has negative dimension of mass
and may therefore be expected to be absolutely convergent.
The following explicit calculation however shows that this
is not so.

The integral over $t$ in eq.~(3.14) can be performed analytically
starting from the time-momentum representation (A.7) of the 
field propagator. After some algebra and the substitution 
$q\to q-\frac{1}{2}p$, the integration leads to the expression
\equation{
   J_1(\om,p)=(2\mu_0-i\om)\int_q
   \left\{2(q^2+\frac{1}{4}p^2+m_0^2)+(2\mu_0-i\om)(4\mu_0-i\om)
   \right\}
   \noenum
   \nexteq{2.0ex}
   \qquad\times\left\{
   4(qp)^2+(2\mu_0-i\om)^2\left[
   4(q^2+\frac{1}{4}p^2+m_0^2)-i\om(4\mu_0-i\om)\right]\right\}^{-1}
   \noenum
   \nexteq{2.5ex}
   \qquad\times\left\{((q+\frac{1}{2}p)^2+m_0^2)((q-\frac{1}{2}p)^2+m_0^2)
   \right\}^{-1}.
   \enum
}
The integrand in this formula is a singularity-free
function of $\om,p$ and $q$, but at large $q$
the integral is logarithmically divergent in four dimensions.
A somewhat lengthy calculation then shows that
\equation{
   J_1(\om,p)\mathrel{\mathop=_{\eps\to0}}
   {1\over16\pi^2\eps}\left\{
   (2\mu_0-i\om)\left(1+
   \sqrt{1+{p^2\over(2\mu_0-i\om)^2}}\right)\right\}^{-1}
   +\rmO(1)
   \enum
}
(see appendix B; the branch of the square root to be taken is 
the principal one).

\section 4. Relation to the ordinary correlation functions

Since the stochastic molecular dynamics simulates the field theory with
action (2.1), the equal-time autocorrelation functions
\equation{
   \tilde{\cal C}_n(p_1,\ldots,p_n)=\int_{\om_1\ldots\om_n}
   \tilde{\cal A}_n(\om_1,p_1;\ldots;\om_n,p_n)
   \enum
}
must coincide with the ordinary correlation functions of the 
fundamental field in momentum space [\ref{Horowitz}]. 
In this section, we show that the 
two- and the four-point autocorrelation functions 
do have this property at one-loop order of 
perturbation theory. Partly the calculation serves as
a consistency check, but some of the intermediate results 
will be helpful in sect.~5 as well, where we discuss the 
non-renormalizability of the stochastic molecular dynamics.

\subsection 4.1 Two-point function

Recalling the results obtained in subsect.~3.2, the 
two-point autocorrelation function is given by
\equation{
   \tilde{\cal A}_2(\om,p;\nu,q)=(2\pi)^{D+1}\delta(\om+\nu)\delta(p+q)
   \noenum
   \nexteq{2.0ex}
   {\phantom{\tilde{\cal A}_2(\om,p;\nu,q)=}}\times
   \left\{\tilde{G}(\om,p)
   +{g_0\over2}I_1{\partial\over\partial m_0^2}\tilde{G}(\om,p)
   +\rmO(g_0^2)\right\}.
   \enum
}
Using the time-momentum representation (A.7) of the field propagator,
the integrals over the frequencies are easily worked out 
and one recovers the familiar expression
\equation{
   \tilde{\cal C}_2(p,q)=(2\pi)^D\delta(p+q)
   \left\{(p^2+m_0^2)^{-1}-{g_0\over2}I_1(p^2+m_0^2)^{-2}
   +\rmO(g_0^2)\right\}
   \enum
}
for the two-point correlation function in the $\phi^4$ theory.

\subsection 4.2 Four-point function at leading order

The leading-order contribution
\equation{
   \tilde{\cal A}_4^{(0)}(\om_1,p_1;\ldots;\om_4,p_4)=
   \raise-0.33cm\hbox{\epsfxsize=5.4cm\epsfbox{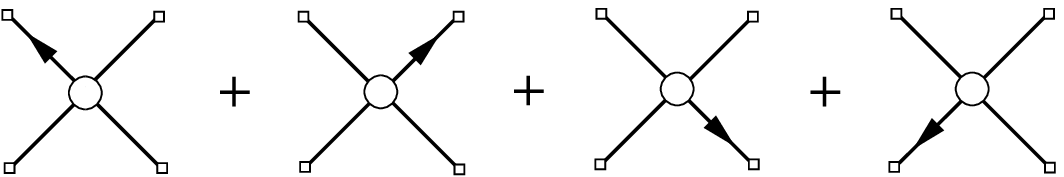}}
   \enum
}
to the four-point autocorrelation function is a sum of four
diagrams. If one substitutes
\equation{
   2\pi\,\delta(\om_1+\ldots+\om_4)=
   \int\rmd t\,\rme^{-it(\om_1+\ldots+\om_4)}
   \enum
}   
for the frequency-conservation $\delta$-function, the integrals over
the frequencies that lead from the autocorrelation to the ordinary
correlation functions factorize and after a few further steps
one obtains
\equation{
   \tilde{\cal C}_4^{(0)}(p_1,\ldots,p_4)=(2\pi)^D\delta(p_1+\ldots+p_4)
   g_0\int_0^{\infty}\rmd t\,
   \partial_t\bigl\{\hat{G}(t,p_1)\ldots\hat{G}(t,p_4)\bigr\}
   \enum
}
for the leading-order four-point correlation function.
Use has here been made of the identity (A.6) and
of the fact that the Green function $\hat{K}(t,p)$ vanishes
at negative times $t$.
Performing the time integration in eq.~(4.6), the correlation
function
\equation{
   \tilde{\cal C}_4^{(0)}(p_1,\ldots,p_4)=(2\pi)^D\delta(p_1+\ldots+p_4)
   (-g_0)(p_1^2+m_0^2)^{-1}\ldots(p_4^2+m_0^2)^{-1}
   \enum
}
is then seen to coincide with the expected expression.

\subsection 4.3 Four-point function at one-loop order

The second-order contribution $\tilde{\cal A}^{(1)}_4(\om_1,p_1;\ldots;\om_4,p_4)$
to the four-point autocorrelation function is a sum of terms proportional
to the integrals $I_1$, $J_0$ and $J_1$. There are 28 diagrams
with an insertion of diagram 1 in one of the external lines.
The sum of all these contributions to the four-point function is
\equation{
   \left.\tilde{\cal A}_4^{(1)}(\om_1,p_1;\ldots;\om_4,p_4)\right|_{I_1}=
   {g_0\over2}I_1{\partial\over\partial m_0^2}
   \tilde{\cal A}_4^{(0)}(\om_1,p_1;\ldots;\om_4,p_4).
   \enum
}
As discussed in subsect.~3.3, diagram 2 is a linear combination
of the integrals $J_0$ and $J_1$, while diagram 3 is
expressed through $J_1$ alone. Collecting all 
terms proportional to $J_0$, their sum is found to be given by
\equation{
   \left.\tilde{\cal A}_4^{(1)}(\om_1,p_1;\ldots;\om_4,p_4)\right|_{J_0}=
   -{g_0\over2}\bigl\{
   J_0(p_1+p_2)+J_0(p_1+p_3)+J_0(p_1+p_4)\bigr\}
   \noenum
   \nexteq{2.0ex}
   {\phantom{\left.\tilde{\cal A}_4^{(1)}(\om_1,p_1;\ldots;\om_4,p_4)\right|_{J_0}=}}
   \times\tilde{\cal A}_4^{(0)}(\om_1,p_1;\ldots;\om_4,p_4).
   \enum
}
Both expressions (4.8) and (4.9) are easily integrated over the frequencies,
because the only factor that depends on $\om_1,\ldots,\om_4$ 
is the tree-level four-point function. One then discovers that 
the expected result for $\tilde{\cal C}_4^{(1)}(p_1,\ldots,p_4)$ is obtained
already from these two contributions to the four-point function.

In order to show that the remaining terms (i.e.~those proportional to $J_1$)
vanish when integrated over the frequencies, first consider the channel 
where the frequency-momentum combination
\equation{
   (\om,p)=(\om_1+\om_2,p_1+p_2)
   \enum
}
flows through the diagrams $2$ and $3$.
Dropping the terms proportional to $J_0$
and using the identity (A.5), one obtains
\equation{
   \raise-0.35cm\hbox{\epsfxsize=1.9cm\epsfbox{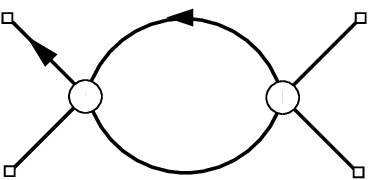}}%
   =-{g_0\over2}J_1(\om,p)\times\Biggl\{\,%
   \raise-0.33cm\hbox{\epsfxsize=5.4cm\epsfbox{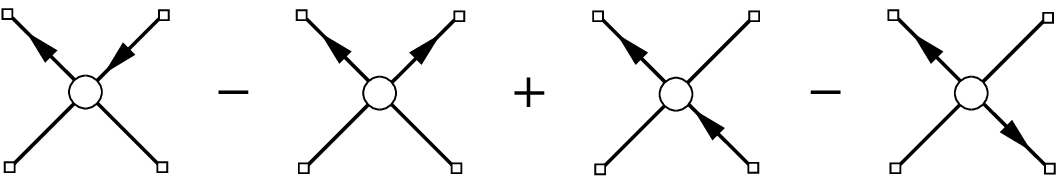}}%
   \,\Biggr\}
   \enum
}
for the contribution of diagram $2$ in this channel.
The contribution of the other diagram is similarly given by
\equation{
   \raise-0.35cm\hbox{\epsfxsize=1.9cm\epsfbox{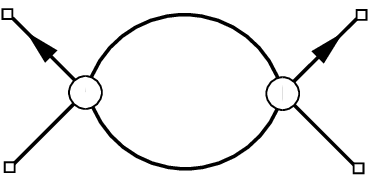}}%
   =-{g_0\over2}\left[J_1(\om,p)+J_1(-\om,p)\right]\times\Biggl\{\,%
   \raise-0.33cm\hbox{\epsfxsize=0.94cm\epsfbox{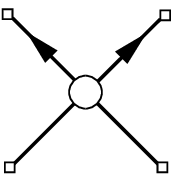}}%
   \,\Biggr\},
   \enum
}
where the prefactor includes the symmetry factor
of the diagram. In total there are six further diagrams that 
differ from the diagrams $2$ and $3$ by a different distribution
of the arrows to the external lines, each of them being given
by the corresponding expression (4.11) or (4.12) with the 
proper assignment of arrows. The
sum of all these contributions to the four-point
function is then equal to
\equation{
   -{g_0\over2}J_1(\om,p)\times\Biggl\{\,%
   \raise-0.33cm\hbox{\epsfxsize=5.4cm\epsfbox{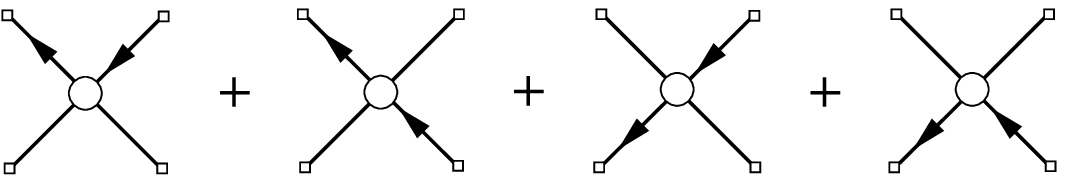}}%
   \,\Biggr\}+\{\om_k\to-\om_k\}
   \enum
}
i.e.~all terms where the arrows are both ingoing 
or both outgoing cancel in the sum.

It is not difficult to show that the terms in eq.~(4.13) vanish
when integrated over the frequencies. The integral over the first term, 
for example, is proportional to
\equation{
   \int_{\om_1,\ldots,\om_4}\!\delta(\om_1+\ldots+\om_4)
   J_1(\om,p)\tilde{K}(\om_1,p_1)
   \tilde{G}(\om_2,p_2)\tilde{G}(\om_3,p_3)\tilde{K}(-\om_4,p_4).
   \enum
}
Eliminating $\om_4$, the integral becomes
\equation{
   \int_{\om_1,\om_2,\om_3}\!
   J_1(\om_1+\om_2,p)\tilde{K}(\om_1,p_1)
   \tilde{G}(\om_2,p_2)\tilde{G}(\om_3,p_3)
   \tilde{K}(\om_1+\om_2+\om_3,p_4),
   \enum
}
and if one first integrates over $\om_1$, the term is seen
to vanish, because the integrand has no singularities 
in the half-plane $\Im \om_1\geq0$
and falls off at least like $\om_1^{-4}$ at large $\om_1$. 
Exactly the same happens for all other terms in eq.~(4.13) and also
those in the other two frequency-momentum channels.

\section 5. Non-renormalizability of the stochastic molecular dynamics

We now address the question whether the 
ultraviolet singularities of the autocorrelation 
functions can be canceled by the addition of 
local counterterms to the evolution equation (2.5).

\subsection 5.1 Parameter renormalization

Evidently, the list of counterterms must include 
those corresponding to the usual parameter and field 
renormalization that is required for the renormalization 
of the ordinary correlation functions. 
In the minimal subtraction scheme,
the bare coupling and mass are 
related to the renormalized parameters $g$ and $m$ through
\equation{
   g_0=M^{2\eps}g\,
   \Bigl\{1+{3g\over32\pi^2\eps}+\rmO(g^2)\Bigr\},
   \enum
   \nexteq{2.5ex}
   m_0^2=m^2\,\Bigl\{1+{g\over32\pi^2\eps}+\rmO(g^2)\Bigr\},
   \enum
}
where $M$ denotes the normalization mass. To one-loop order,
the fundamental field does not need to be renormalized in this theory.

Recalling eqs.~(4.2), (4.8) and (4.9), it is then
immediately clear that the parameter renormalization 
cancels the singularities 
of the two- and four-point autocorrelation functions
which derive from the poles (3.8) and (3.16) of 
the integrals $I_1$ and $J_0$.

\subsection 5.2 Non-renormalizability of the four-point function

The four-point autocorrelation function has further singularities
proportional to the divergent part (3.18) of the integral $J_1$.
As explained in sect.~4, the ordinary four-point correlation function
does not receive any contributions from this integral (and
is therefore finite after the parameter renormalization),
but the terms proportional to $J_1$ do contribute to the 
autocorrelation function at non-zero time separations.

The residue of the pole in eq.~(3.18) is the Fourier transform of 
a distribution
\equation{
   {\rme^{-2\mu_0t}\over32\pi^4x^2}\theta(t)\delta(t^2-x^2)
   \enum
}
supported on the light cone $t=|x|$. 
Both diagrams 2 and 3 thus
have a non-local singularity that cannot be canceled by
including local counterterms in the stochastic molecular
dynamics. The latter is therefore not renormalizable.

The presence of the singularity (5.3) can be understood 
by noting that the integrand of the integral
\equation{
   J_1(\om,p)=\int_0^{\infty}\rmd t\int\rmd^D\kern-1pt x\,\rme^{i\om t-ipx}
   G(t,x)^2
   \enum
}
has a non-integrable singularity in $D=4$ dimensions
proportional to $(t-|x|)^{-1}$ (see subsect.~A.3).
Such light-cone singularities are a characteristic 
feature of Green functions of hyperbolic wave
equations and the non-renormalizability of 
the stochastic molecular dynamics is thus seen to be related to
its hyperbolic nature.

In ordinary field theory, one-loop integrals do not have
non-local ultraviolet singularities,
because they can be Wick rotated to Euclidean space
where the propagators are singular at coinciding points only.
The spectral condition and the locality of the theory guarantee 
that no singularities stand in the way of the Wick rotation [\ref{BS}].
In the case of the diagrams 2 and 3, however,
the integrands have poles in all quadrants 
of the complex frequency plane and the integrals (3.9) and (3.10)
consequently cannot be Wick rotated 
without generating additional terms.

\topinsert
\vbox{
\vskip0.0cm
\centerline{\epsfxsize=7.4cm\epsfbox{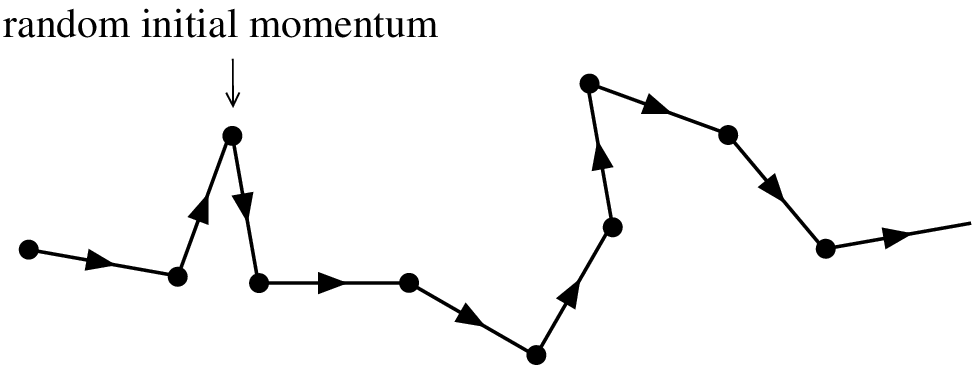}}
\vskip0.3cm
\figurecaption{%
The HMC algorithm moves the fundamental field $\phi$ through field space
along a piecewise smooth curve. In the smooth segments of the curve,
the field is evolved from time $t=0$ to some time $t=\tau$ 
according to the molecular-dynamics
equations, starting from the current field $\phi$ 
and Gaussian random values for its momentum $\pi$.
}
}
\endinsert

\subsection 5.3 Implications for the HMC algorithm

In practice, the HMC algorithm involves a numerical integration
of the (ordinary) molecular-dynamics equations and 
acceptance-rejection steps to correct for the integration errors.
For simplicity the integration is assumed to be exact in this 
section. No acceptance-rejection steps are then required and 
whether one uses the first- or the second-order form of the 
molecular-dynamics equations makes no difference.

The molecular-dynamics trajectories 
generated by the algorithm are smooth segments 
of a continuous curve in field space (see fig.~3).
Along the trajectories, the $n$-point autocorrelation functions
in the time-momentum representation,
\equation{
   \hat{\cal A}_n(t_1,p_1;\ldots;t_n,p_n)=
   \langle\hat{\phi}(t_1,p_1)\ldots\hat{\phi}(t_n,p_n)\rangle,
   \qquad 0\leq t_k\leq\tau,
   \enum
}
may be defined, where
the bracket $\langle\ldots\rangle$ stands for the
average over all trajectories in an infinitely long simulation.
The autocorrelation functions (5.5) only
describe the dynamical properties of the algorithm in
the specified range of times,
but the discussion in the following paragraphs
shows that already these correlation functions are 
not renormalizable.

The average over trajectories in eq.~(5.5) amounts to 
taking the average over the initial values of the 
field $\phi$ and its momentum $\pi=\partial_t\phi$.
Since these are distributed according to the equilibrium
distribution (a Gaussian in the case of the momentum), 
the average coincides with the ordinary expectation value. 
In perturbation theory,
the correlation functions can therefore be calculated by solving 
the (non-stochastic) molecular-dynamics equations in the 
range $0\leq t\leq\tau$ with prescribed initial data 
at $t=0$ and by computing the expectation value of the product
$\hat{\phi}(t_1,p_1)\ldots\hat{\phi}(t_n,p_n)$ using
the standard Feynman rules for the correlation functions of 
the initial data.

In the case of the stochastic molecular dynamics,
the computation of the autocorrelation functions in the 
time-momentum representation can be organized in
the same way. A notable difference is that 
the contractions of the noise field give rise to additional diagrams,
but since all these diagrams disappear in the limit $\mu_0\to0$,
it is clear that the autocorrelation functions (5.5) 
are given by 
\equation{
   \hat{\cal A}_n(t_1,p_1;\ldots;t_n,p_n)=
   \lim_{\mu_0\downarrow0}
   \int_{\om_1,\ldots,\om_n}
   \rme^{-i(\om_1t_1+\ldots+\om_nt_n)}
   \tilde{\cal A}_n(\om_1,p_1;\ldots;\om_n,p_n),
   \noenum
   \nexteq{-0.3ex}
   \enum
}
where the autocorrelation functions on the right are those
discussed in the previous sections. Note that the frequency integrals
must be performed before
$\mu_0$ is taken to zero, as otherwise one may run into 
infrared-singular intermediate expressions.

In view of its relation to the stochastic molecular dynamics,
as expressed through eq.~(5.6), and since the distribution (5.3) 
remains non-local at $\mu_0=0$, we are thus led to conclude 
that also the HMC algorithm is not renormalizable.

\subsection 5.4 The Langevin limit 

As already mentioned in subsect.~2.1, the stochastic molecular-dynamics
equation (2.5) is equivalent to the Langevin equation in the 
limit $\mu_0\to\infty$ up to a rescaling of the simulation
time\kern1pt\footnote{$\dagger$}{\footnotefont%
In lattice field theory, the Langevin limit can also be
reached together with the continuum limit 
by setting $\mu_0$ to some fixed value in 
units of the lattice spacing.}.
The associated $n$-point autocorrelation functions,
\equation{
   \tilde{\cal A}_n^{\infty}(\om_1,p_1;\ldots;\om_n,p_n)
   =\lim_{\mu_0\to\infty}(2\mu_0)^{-n}
   \tilde{\cal A}_n(\om_1/2\mu_0,p_1;\ldots;\om_n/2\mu_0,p_n),
   \enum
}
are known to be renormalizable to all orders
of perturbation theory [\ref{Zinn}]. 

It may be instructive to see how exactly the renormalizability
of the autocorrelation functions gets restored at one-loop
order when $\mu_0$ is sent to infinity. 
To this end, first note that the Green function
\equation{
   \lim_{\mu_0\to\infty}\tilde{K}(\om/2\mu_0,p)=
   {1\over-i\om+p^2+m_0^2}
   \enum
}
assumes the expected form, which is smooth in position space
except for a singularity at the origin.
At one-loop order, the limit of 
the two-point function and the renormalizable parts of
the four-point function is then easily determined 
starting from the identities (4.2), (4.8) and (4.9). 
These remain valid in the limit and only
the tree-level autocorrelation functions are replaced by 
the corresponding expressions involving the propagator
\equation{
   \lim_{\mu_0\to\infty}(2\mu_0)^{-1}
   \tilde{G}(\om/2\mu_0,p)={2\over\om^2+(p^2+m_0^2)^2}
   \enum
}
and the Green function (5.8).

The contribution to the four-point function proportional
to the integral $J_1$ (which contains
the non-removable ultraviolet singularity
in the case of the stochastic molecular dynamics)
however changes its character, because
the integral
\equation{
   \lim_{\mu_0\to\infty}(2\mu_0)^{-1}J_1(\om/2\mu_0,p)=
   \int_q\left\{-i\om+2q^2+\frac{1}{2}p^2+2m_0^2\right\}^{-1}
   \noenum
   \nexteq{2.5ex}
   \qquad\times\left\{((q+\frac{1}{2}p)^2+m_0^2)((q-\frac{1}{2}p)^2+m_0^2)
   \right\}^{-1}
   \enum
}
turns out to be absolutely convergent.
In the Langevin limit, all ultraviolet singularities at
one-loop order are thus canceled by the parameter renormalization, 
as is expected to be the case in this theory [\ref{Zinn}].

\section 6. Concluding remarks

The HMC algorithm is currently the
preferred simulation algorithm in lattice QCD.
In the past two decades,
various improvements were included in this algorithm,
many of them with the aim of reducing the computational
effort required at small sea-quark masses (see ref.~[\ref{LesHouches}]
for a recent review). Its scaling behaviour with respect to 
the lattice spacing has not received as much attention
so far, but rapidly becomes an important issue when the
continuum limit is approached.

While the dynamical properties of the HMC algorithm are well
understood in free field theory [\ref{GHMC}], the
situation in the presence of interactions tends to 
be rather more complicated. In particular, certain lattice artifacts
(topology-changing tunneling transitions, for example,
or unphysical critical points in the space of bare couplings)
can cause large autocorrelations. 
The results obtained in this paper show
that even in the absence of such effects
there is no reason to expect that the HMC algorithm scales 
essentially as in a theory of free fields.
Evidently, the non-renormalizability
of the algorithm does not imply that it is invalid
or unusable close to the continuum limit, 
but without further insight its scaling behaviour 
is unpredictable in interacting theories.

The HMC algorithm and the stochastic molecular dynamics 
may conceivably fall into the universality class of the 
Langevin equation. Independently of whether this is the case
or not, it may be worth looking for renormalizable algorithms
where the simulation time has scaling dimension 
less than $2$.
Eventually such algorithms might turn out to be
faster than the HMC algorithm and they would have the
advantage that their efficiency
at small lattice spacings is predictable.

\appendix A. Properties of the Green function 
             ${\mib K}$(${\mib t,x}$)

\vskip-2.5ex

\subsection A.1 Definition

The Fourier transform (2.9) of the Green function
is a smooth function of $\om$ and $p$ that satisfies
\equation{
   \bigl|\tilde{K}(\om,p)\bigr|^2\leq 
   C\left(\om^2+p^2+m_0^2\right)^{-1}
   \enum
}
for some (mass-dependent) constant $C$.
$\tilde{K}(\om,p)$ is therefore a tempered distribution and so is 
the Green function in position space. However, the Fourier 
integral (2.8) is not absolutely convergent and is
to be understood in the sense of distributions.
All these comments also apply to the propagator (3.3) 
of the basic field.

As a distribution, $K(t,x)$ satisfies the wave equation
\equation{
   {\cal D}K(t,x)=\delta(t)\delta(x),
   \enum
}
where ${\cal D}$ is given by eq.~(2.7). 
Since the polynomial representing 
${\cal D}$ in frequency-momentum space is nowhere equal to
zero, the Green function is the unique tempered distribution 
that solves eq.~(A.2). In particular, one does not
have the freedom of specifying retarded or advanced boundary 
conditions.

\subsection A.2 Time-momentum representation

Using the residue theorem, it is possible to work out the Green 
function
\equation{
   \hat{K}(t,p)=\int_{\om}\rme^{-i\om t}\tilde{K}(\om,p)
   \enum
   \nexteq{2.0ex}
   {\phantom{\hat{K}(t,p)}}=
   \theta(t)\rme^{-\mu_0t}{\sin(\eps_pt)\over\eps_p},
   \qquad
   \eps_p=\left(p^2+m_0^2-\mu_0^2\right)^{1/2}.
   \enum
}
in the time-momentum representation. Note that $\eps_p$ is imaginary
at small momenta if $\mu_0>m_0$, but the Green function 
always decays exponentially in the time direction. 

In the case of the field propagator, the identity
\equation{
   i\om\tilde{G}(\om,p)=
   \tilde{K}(\om,p)-\tilde{K}(-\om,p)
   \enum
}
and thus
\equation{
   \partial_t\hat{G}(t,p)=\hat{K}(-t,p)-\hat{K}(t,p)
   \enum
}
may be used to show that
\equation{
   \hat{G}(t,p)={\rme^{-\mu_0|t|}\over p^2+m_0^2}
   \left\{\cos(\eps_pt)+\mu_0{\sin(\eps_p|t|)\over\eps_p}\right\}.
   \enum
}
An immediate consequence of these results is
that $K(t,x)$ and $G(t,x)$ become smooth functions of $t$
at all $t\neq0$ when smeared with a test function in $x$.
Moreover, 
\equation{
   K(t,x)\to0, \qquad\partial_tK(t,x)\to\delta(x),
   \enum
}
as $t\downarrow0$.

\subsection A.3 Explicit expression for $K(t,x)$ in four dimensions

In dimension $D=4$,
the Green function in position space, $K(t,x)$, 
can be calculated analytically.
While the expression is of some interest in the context of
our discussion of the non-renormalizability
of the stochastic molecular dynamics in sect.~5, 
its exact form is not needed and the proof of the 
results quoted below is therefore omitted.

First note that the product $\rme^{t\mu_0}K(t,x)$ 
is invariant under Lorentz transformations in $5$ dimensions.
As a consequence, and since the Green function vanishes at negative times, 
its support must be contained in the forward light cone $t\geq|x|$.
Inside the light cone, i.e.~at all $t>|x|$, the formula
\equation{
   K(t,x)=-{\rme^{-\mu_0t}\over4\pi^2}\left\{
   {\cos(\eps_0 s)\over s^3}+
   {\eps_0\sin(\eps_0s)\over s^2}\right\}_{s=(t^2-x^2)^{1/2}}
   \enum
}
holds, where $\eps_0=(m_0^2-\mu_0^2)^{1/2}$. 
While the Green function has no singularities there,
the formula shows that it diverges proportionally to 
\equation{
  \rme^{-\mu_0t}(t^2-x^2)^{-3/2}
  \enum
}
along the light cone $t=|x|$.

\appendix B. Proof of eq.~(3.18)

First note that the integral
\equation{
   \hat{J}_1(\om,p)={1\over2}\int_q
   {2\mu_0-i\om\over
   \left[(qp)^2+(2\mu_0-i\om)^2(q^2+m_0^2)\right]
   (q^2+m_0^2)}
   \enum
}
has the same divergent part as $J_1$,
because the integrand of the difference $\hat{J}_1-J_1$
falls off like $(q^2)^{-3}$ at large $q$ and is therefore 
absolutely integrable in four dimensions. 

In view of the reality property
\equation{
   \hat{J}_1(-\om,p)=\hat{J}_1(\om,p)^{\ast},
   \enum
}
it suffices to calculate $\hat{J}_1(\om,p)$ at all positive $\om$.
The imaginary part of the factor
\equation{
   (qp)^2+(2\mu_0-i\om)^2(q^2+m_0^2)
   \enum
}
is then strictly negative and the Feynman parameter representation
\equation{
   \hat{J}_1(\om,p)={i\over2}(2\mu_0-i\om)
   \int_0^{\infty}\rmd u\rmd v
   \int_q\rme^{-iu
   \left[(qp)^2+(2\mu_0-i\om)^2(q^2+m_0^2)\right]}\,
   \rme^{-v(q^2+m_0^2)}
   \noenum
   \nexteq{-1.0ex}
   \enum
}
is therefore well defined.

The Gaussian integral over the momentum $q$ can now be performed and
leads to the formula
\equation{
   \hat{J}_1(\om,p)={i\over2(4\pi)^{D/2}}(2\mu_0-i\om)
   \int_0^{\infty}\rmd u\rmd v\,
   \rme^{-(uz+v)m_0^2}
   \noenum
   \nexteq{2.0ex}
   {\phantom{\hat{J}_1(\om,p)=}}
   \times
   \left\{u(z+ip^2)+v\right\}^{-1/2}
   \left\{uz+v\right\}^{-(D-1)/2},
   \enum
}
where 
\equation{
   z=i(2\mu_0-i\om)^2=4\mu_0\om+i(4\mu_0^2-\om^2).
   \enum
}
In eq.~(B.5), the phases of the expressions in the curly brackets
range from $-{1\over2}\pi$ to $+{1\over2}\pi$ and 
it is understood that the corresponding branch of 
their powers is taken.

The integral (B.5) is absolutely convergent for $D<4$
and a holomorphic function of $z$ in the half-plane $\Re z>0$.
Its values along the real axis $z>0$ can be computed
by first substituting $u\to u/z$ and subsequently
\equation{
   u=ts,\quad v=t(1-s),
   \enum
   \nexteq{2.5ex}
   0\leq t<\infty,\quad 0\leq s\leq1.
   \enum
}
The integral then factorizes into analytically calculable
integrals and one finds that it is given by
\equation{
   {i\Gamma(\eps)\over m_0^{2\eps}(4\pi)^{D/2}}
   {2\mu_0-i\om\over z}\left(
   1+\sqrt{1+i{p^2\over z}}\right)^{-1}.
   \enum
}
This expression analytically extends
to the half-plane $\Re z>0$
and therefore coincides with the integral (B.5) at all these
values of $z$. Inserting eq.~(B.6), the result 
\equation{
   \hat{J}_1(\om,p)={\Gamma(\eps)\over m_0^{2\eps}(4\pi)^{D/2}}
   \left\{
   (2\mu_0-i\om)\left(1+
   \sqrt{1+{p^2\over(2\mu_0-i\om)^2}}\right)\right\}^{-1}
   \enum
}
is thus obtained,
where the branch of the square root 
with positive real part is to be taken. 
This formula holds for both positive and negative $\om$.

\beginbibliography


\bibitem{Zinn}
J. Zinn--Justin,
{\it Renormalization and stochastic quantization},
Nucl. Phys. B275 [FS17] (1986) 135

\bibitem{ZinnZwanziger}
J. Zinn--Justin, D. Zwanziger, 
{\it Ward identities for the stochastic quantization of gauge fields},
Nucl. Phys. B295 [FS21] (1988) 297


\bibitem{MunozAlvarez}
A. Mu\~{n}oz Sudupe, R. F. Alvarez-Estrada,
{\it Renormalization constants for the propagator of the stochastically
quantized Yang-Mills field theory},
Phys. Lett. B164 (1985) 102;
{\it Beta function for Yang-Mills field theory in stochastic quantization},
Phys. Lett. B166 (1986) 186

\bibitem{Okano}
K. Okano,
{\it Background field method in stochastic quantization},
Nucl. Phys. B289 (1987) 109


\bibitem{BaulieuZwanziger}
L. Baulieu, D. Zwanziger,
{\it ${\it QCD}_4$ from a five-dimensional point of view},
Nucl. Phys. B581 (2000) 604


\bibitem{DelDebbioTauQ}
L. Del Debbio, H. Panagopoulos, E. Vicari,
{\it $\theta$-dependence of SU(N) gauge theories},
JHEP 08 (2002) 044


\bibitem{SchaeferTauQ}
S. Schaefer, R. Sommer, F. Virotta,
{\it Investigating the critical slowing down of QCD simulations},
PoS(LAT2009)032;
{\it Critical slowing down and error analysis in lattice QCD simulations},
Nucl. Phys. B845 (2011) 93


\bibitem{Villasimius}
M. L\"uscher,
{\it Topology, the Wilson flow and the HMC algorithm},\hfill\break
PoS(Lattice 2010)015


\bibitem{SommerScale}
R. Sommer,
{\it A new way to set the energy scale in lattice gauge theories
and its applications to the static force and $\alpha_s$ in
SU(2) Yang--Mills theory},
Nucl. Phys. B411 (1994) 839


\bibitem{HMC}
S. Duane, A. D. Kennedy, B. J. Pendleton, D. Roweth,
{\it Hybrid Monte Carlo},
Phys. Lett. B195 (1987) 216.


\bibitem{GHMC}
A. D. Kennedy, B. Pendleton,
{\it Cost of the generalized Hybrid Monte Carlo algorithm for 
free field theory},
Nucl. Phys. B607 (2001) 456


\bibitem{Horowitz}
A. M. Horowitz,
{\it Stochastic quantization in phase space},
Phys. Lett. 156B (1985) 89;
{\it The second order Langevin equation and numerical simulations},
Nucl. Phys. B280 [FS18] (1987) 510;
{\it A generalized guided Monte Carlo algorithm},
Phys. Lett. B268 (1991) 247


\bibitem{BeccariaCurci}
M. Beccaria, G. Curci,
{\it The Kramers equation simulation algorithm. 1. Operator analysis},
Phys. Rev. D49 (1994) 2578

\bibitem{JansenLiu}
K. Jansen, C. Liu,
{\it Kramers equation algorithm for simulations of QCD
with two flavors of Wilson fermions and gauge group SU(2)},
Nucl. Phys. B453 (1995) 375 [E: {\it ibid.} B459 (1996) 437]


\bibitem{BS}
N. N. Bogoliubov, D. V. Shirkov,
{\it Introduction to the theory of quantized fields},
Wiley--Interscience, New York, 1959


\bibitem{LesHouches}
M.~L\"uscher,
{\it Computational strategies in lattice QCD},
Lectures given at the
Summer School on ``Modern perspectives in lattice QCD'',
Les Houches, August 3-28, 2009,
arXiv:1002.4232 [hep-lat]

\endbibliography

\bye